\documentclass[12pt]{article}
\usepackage{amssymb,amsmath,epsfig}
\begin{document}
\parindent=0pt
\parskip=6pt
\rm

\vspace*{0.5cm}

\begin{center}

\normalsize {\bf THERMODYNAMIC PROPERTIES OF THE PHASE TRANSITIONS IN A
CLASS OF SPIN-TRIPLET FERROMAGNETIC SUPERCONDUCTORS}

\vspace{0.3cm}

{\bf D. V. Shopova, D. I. Uzunov}

CPCM Laboratory, Institute of Solid State Physics,
 Bulgarian Academy of Sciences, BG-1784 Sofia, Bulgaria.\\
emails: sho@issp.bas.bg, uzun@issp.bas.bg
\end{center}

\vspace{0.3cm}


\vspace{0.3cm}

\begin{abstract}
Magnetic susceptibility, entropy and specific heat are calculated at
the equilibrium points of phase transition to a phase of coexistence of
ferromagnetic order and superconductivity in a new class of
spin-triplet ferromagnetic superconductors. The results are discussed
in view of application to metallic ferromagnets as UGe$_2$, ZrZn$_2$,
URhGe.
\end{abstract}

\vspace{0.5cm}

{\bf 1. Introduction} 

Recently, a spin-triplet superconducting phase has
been discovered in a class of new ferromagnetic superconductors
(UGe$_2$~[1], ZrZn$_2$~[2], URhGe~[3]). In these metallic compounds the
superconductivity exists only in the domain of stability of the
ferromagnetic order. Moreover, the ferromagnetic order enhances the
superconductivity because of the unconventional Cooper pairing with
spin S =1~[4-7]. A phenomenological model~[7] of Ginzburg--Landau (GL)
type has been used to describe the possible ordered phases and the
phase diagram of these metallic compounds~[8,9].

In this paper we present the results of our calculation of magnetic
susceptibility, entropy and specific heat near the phase transitions in
ferromagnetic superconductors with a spin-triplet superconductivity.
For our aims we shall use the GL free energy introduced in Ref.~[7] and
results published in Refs.~[8,9].

{\bf 2. Model and phase diagram}

 We consider the GL free energy~[7-9]
$F=\int d^3 x f(\psi, \vec{{\cal{M}}})$, where
\begin{equation}
\label{eq1} f = \frac{\hbar^2}{4m} (D_j\psi)^{\ast}(D_j\psi) +
a_s|\psi|^2 + \frac{b}{2}|\psi|^4 + a_f\vec{{\cal{M}}}^2 +
\frac{\beta}{2}{\cal{\vec{M}}}^4 + i\gamma_0
\vec{{\cal{M}}}.(\psi\times \psi^*)\;.
\end{equation}
In Eq.~(\ref{eq1}), $D_j =(\nabla - 2ieA_j/\hbar c)$, and $A_j$ ($j =
1,2,3$) are the components of the vector potential $\vec{A}$ related
with the magnetic induction $\vec{B} = \nabla \times \vec{A}$. The
complex vector $\psi = (\psi_1,\psi_2,\psi_3) \equiv \{\psi_j\}$ is the
superconducting order parameter, corresponding to the spin-triplet
Cooper pairing and $\vec{{\cal{M}}}= \{{\cal{M}}_j\}$ is the
magnetization. The coupling constant $\gamma_0 = 4\pi J>0$ is given by
the ferromagnetic exchange parameter $(J>0)$. Coefficients $a_s =
\alpha_s(T-T_s)$ and $a_f = \alpha_f(T-T_f)$ are expressed by the
positive parameters $\alpha_s$ and $\alpha_f$ as well as by the
superconducting $(T_s)$ and ferromagnetic ($T_f$) critical temperatures
in the decoupled case, when ${\cal{M}}\psi_i\psi_j$-interaction is
ignored; $b > 0$ and $\beta > 0$ as usual.

We assume that the magnetization ${\cal{M}}$ is uniform, which is a
reliable assumption outside a quite close vicinity of the magnetic
phase transition but keep the spatial ($\vec{x}-$) dependence of
$\psi$. The reason is that the relevant dependence of $\psi$ on
$\vec{x}$ is generated by the diamagnetic effects arising from the
presence of ${\cal{M}}$ and the external magnetic field $\vec{H}$~[8,9]
rather than from fluctuations of $ \psi $ (this effect is extremely
small and can be ignored). First term in~(\ref{eq1}) will be still
present even for $\vec{H} = 0$ because of the diamagnetic effect
created by the magnetization $\vec{{\cal{M}}} = \vec{B}/4\pi > 0$. As
we shall investigate the conditions for the occurrence of the Meissner
phase where $\psi$ is uniform, the spatial dependence of $\psi$ and,
hence, the first term in r.h.s. of~(\ref{eq1}) will be neglected. This
approximation should be valid when the lower critical field $H_{c1}(T)$
is greater than the equilibrium value of the magnetization ${\cal{M}}$
in the phase of coexistence of superconductivity and ferromagnetism.

We take an advantage of the symmetry of  model~(\ref{eq1}) and avoid
the consideration of equivalent thermodynamic states that occur as a
result of the respective continuous symmetry breaking at the phase
transition point but have no effect on thermodynamics of the system.
That is why we shall assume for concreteness of our calculation that
the magnetization vector is along the $z$-axis: $\vec{{\cal{M}}} =
(0,0,{\cal{M}})$, where ${\cal{M}}\geq 0$. This concrete choice of the
direction of the magnetization vector does not restrict the generality
of the present analysis and leads to the same structure of the ordered
phases as previously predicted and discussed on the basis of general
symmetry group considerations in Ref.~[7].

\begin{figure}
\begin{center}
\epsfig{file=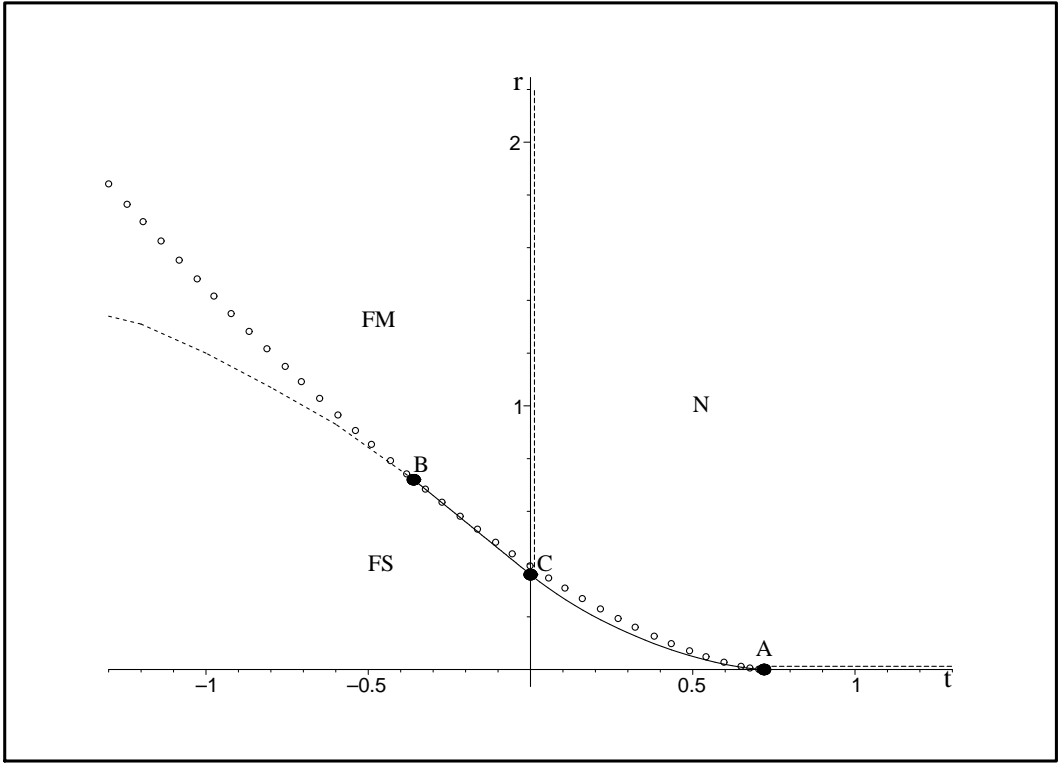,angle=-90, width=14cm}\\
\end{center}
\caption{The phase diagram in the plane ($t$, $r$) with two tricritical
points ($A$ and $B$) and a triple point $C$; $\gamma = 1.2$. Second
order transition lines are dashed, first order transition line is
solid.}
\label{su0903f.fig}
\end{figure}

We find convenient to use the following notations: $\varphi_j
=b^{1/4}\psi_j$, $\varphi_j = \phi_j\mbox{exp}(\theta_j)$, $M =
\beta^{1/4}{\cal{M}}$, $\gamma= \gamma_0/ (b^2\beta)^{1/4}$, $r =
a_s/\sqrt{b}$, $t = a_f/\sqrt{\beta}$, $\phi^2 = (\phi_1^2 + \phi_2^2 +
\phi_3^2)$, and $\theta = (\theta_2-\theta_1)$. We consider the stable
homogeneous phases in a zero external magnetic field ($\vec{H} = 0$)
that are described by uniform order parameters ${\cal{M}}$ and $\psi$.
These phases are three: (i) the normal phase ($\phi = M = 0$),
hereafter referred  as ``N'', (ii) the ferromagnetic phase ($\phi = 0,
M
> 0$), hereafter referred as ``FM'', and (iii) the phase of coexistence of
superconductivity and ferromagnetism $(\phi
> 0, M > 0)$, which will be called ``FS''-phase. The standard N-FM phase
transition occurs at $T_f$. The N-FM, N-FS and FM-FS phase transitions
are shown in Fig.~1~[8,9]. The r-axis above the point $C$ coincides
with the line of the N-FM phase transition of second order. The dashed
lines correspond to phase transitions of second order: FM-FS transition
for $t < (-\gamma^2/4)$ and a N-FS transition for $t > \gamma^2/2$. The
solid lines $BC$ and $CA$ describe first order FM-FS and N-FS
transitions, respectively. The line of circles $BCA$ defines the
borderline of stability and existence of FS. In the domain confined
between the line of circles and the dashed line on the left of the
point $B$ the stability condition for FS is satisfied but the existence
condition is broken. The phase diagram contains two tricritical points
($A$ and $B$) and a triple point ($C$). In the domain between the solid
lines $BC$ and $CA$ and the line of circles $BCA$, FS can be
overheated. Domains of overcooling of phases FM and N below the solid
lines $BC$ and $CA$ also exist~[8,9]. For example, the domain below the
solid line $CA$ and above the positive axes is a region of overcooling
of the N-phase.

{\bf 3. Magnetic susceptibility} 

The magnetic susceptibility $\chi =
(\tilde{\chi}/V) = d^2f(M)/dM^2$ per unit volume ($V$) in a zero
external magnetic field is obtained in the general form
\begin{equation}
\label{eq2} \chi^{-1} = 2t + 6M^2 -
\left(\frac{\partial \phi^2}{\partial M}\right)^2,\;
\end{equation}
where the equilibrium magnetization $M$ should be taken for the
respective equilibrium phase: (a) $M=0$ in the N-phase, (b) $M =
\sqrt{|t|}$ in FM, and (c) in FS, $M$ is given as the maximal
nonnegative root of the equation~[8,9].
\begin{equation}
\label{eq3}
 \frac{\gamma r}{2} = \left(\frac{\gamma^2}{2}-t \right)M -
M^3\:.
\end{equation}
We shall compare the known paramagnetic $(\chi_P = 1/2t; t>0)$ and the
ferromagnetic $(\chi_F = 1/4|t|; t <0)$ susceptibilities with that
corresponding to FS. Note, that both $\chi_P$ and $\chi_F$ are
divergent on the N-FM phase transition line. While the magnetic
susceptibility in phases N and FM and the behaviour of the same
quantity at the N-FM phase transition do not depart from the standard
predictions~[6] we shall see that the phase transitions from FS to N or
FM show quite special magnetic properties. As $\chi$ cannot be
analytically calculated for the whole domain of stability of FS, we
shall consider the close vicinity of the N-FS and FM-FS phase
transition lines.

Near the second order phase transition line on the left of the point
$B$ ($t < -\gamma^2/4$), the magnetization has a smooth behaviour and
the magnetic susceptibility does not exhibit any singularities (jump or
divergence). For $t > \gamma^2/2$, the magnetization is given by $M =
(s_- + s_+)$, where
\begin{equation}
\label{eq4}
 s_{\pm} =\left\{- \frac{\gamma r}{4} \pm \left[
\frac{(t-\gamma^2/2)^3}{27} + \left( \frac{\gamma
r}{4}\right)^2\right]^{1/2} \right\}^{1/3}\;.
\end{equation}
For $r = 0$, $M = 0$, whereas for $|\gamma r| \ll (t - \gamma^2/2)$ and
$r=0$ one may obtain $M \approx -\gamma r/ (2t-\gamma^2) \ll 2t$. This
means that in a close vicinity $(r < 0)$ of $r = 0$ along the second
order phase transition line $(r=0, t>\gamma^2)$ the magnetic
susceptibility is well described by the paramagnetic law $\chi_P =
(1/2t)$. For $r< 0$ and $t \rightarrow \gamma^2/2$, we obtain $M =
-(\gamma r/2)^{1/3}$ which yields
\begin{equation}
\label{eq5}
 \chi^{-1} =  6\left(\frac{\gamma |r|}{2}\right)^{2/3}\:.
\end{equation}

On the phase transition line $AC$ we have ~[8,9]
\begin{equation}
\label{eq6} M_{eq}(t) = \frac{1}{2\sqrt{2}}\left[\gamma^2 - 8t +
\gamma\left(\gamma^2 + 16t \right)^{1/2}\right]^{1/2}
\end{equation}
and, hence,
\begin{equation}
\label{eq7}
 \chi^{-1} = -4t - \frac{\gamma^2}{4}\left[1-3\left( 1 +
 \frac{16t}{\gamma^2}\right)^{1/2}\right]\:.
\end{equation}
At the tricritical point $A$ this result yields $\chi^{-1}(A) = 0$,
whereas at the triple point $C$ with coordinates ($0$, $\gamma^2/4$) we
have $\chi(C) = (2/\gamma^2)$. On the line $BC$ we obtain
$M=\gamma/2$~[8,9] and, hence,
\begin{equation}
\label{eq8} \chi^{-1} = 2t + \frac{\gamma^2}{2}\:.
\end{equation}
At the tricritical point $B$ with coordinates ($-\gamma^2/4$,
$\gamma^2/2$) this result yields $\chi^{-1}(B)= 0$.

The comparison of Eqs.~(7) and (8) with $\chi_P$ corresponding to the
N-phase and with $\chi_F$ corresponding to FM  shows that the magnetic
susceptibility undergoes finite jumps at the first order N-FS and FM-FS
transitions. The finite jumps $\Delta\chi_P$ and $\Delta\chi_F$ can be
easily calculated with the help of $\Delta\chi_{P,F} = (\chi -
\chi_{P,F})$ and Eqs.~(7) and (8). In particular, the jump of $\chi$ at
the triple point $C$ is infinite because of the divergency of $\chi_P$
in the limit $t \rightarrow 0^+$.

{\bf 4. Entropy and specific heat} 

The entropy $S(T) \equiv (\tilde{S}/V)
= -V\partial (f/\partial T) $ and the specific heat $C(T) \equiv
(\tilde{C}/V) = T(\partial S/\partial T)$ per unit volume $V$ are
calculated in a standard way~[6]. We are interested in the jumps of
these quantities on the N-FM, FM-FS, and N-FS transition lines. The
behaviour of $S(T)$ and $C(T)$ near the N-FM phase transition and near
the FM-FS phase transition line of second order on the left of the
point $B$ (Fig.~1) is known from the standard theory of critical
phenomena (see, e.g., Ref.~[6] and for this reason we focus our
attention on the phase transitions of type FS-FM and FS-N for
$(t>-\gamma^2/4$), i.e., on the right of the point $B$ in Fig.~1.

Using the equations for the order parameters $\psi$ and $M$~[8,9] and
applying the standard procedure for the calculation of $S$, we obtain
the general expression
\begin{equation}
\label{eq9}
 S(T) = - \frac{\alpha_s}{\sqrt{b}}\phi^2 -
\frac{\alpha_f}{\sqrt{\beta}}M^2\:.
\end{equation}
The next step is to calculate the entropies $S_{FS}(T)$ and $S_{FM}$ of
the ordered phases FS and FM. Note, that we work with the usual
convention $F_N = Vf_N=0$ for the free energy of the N-phase and,
hence, we must set $S_{N}(T)=0$.

Consider the second order phase transition line ($r=0$,
$t>\gamma^2/2$). Near this line $S_{FS}(T)$ is a smooth function of $T$
and has no jump but the specific heat $C_{FS}$ has a jump at $T=T_s$,
i.e. for $r=0$. This jump is given by
\begin{equation}
\label{eq10}
 \Delta C_{FS}(T_s) = \frac{\alpha_s^2T_s}{b}\left[ 1 -
 \frac{1}{1 - 2t(T_s)/\gamma^2}\right]\:.
\end{equation}
The jump $\Delta C_{FS}(T_s)$ is higher than the usual jump $\Delta
C(T_c) = T_c\alpha^2/b$ known from the Landau theory of standard second
order phase transitions~[6].

The entropy jump $\Delta S_{AC}(T) \equiv S_{FS}(T) $ on the line $AC$
is obtained in the form
\begin{equation}
\label{eq11}
 \Delta S_{AC}(T) = -M_{eq}\left\{\frac{\alpha_s\gamma}{4\sqrt{b}}\left[1
 + \left(1 + \frac{16t}{\gamma^2}\right)^{1/2}\right] -
 \frac{\alpha_f}{\sqrt{\beta}}M_{eq}\right\}\:,
\end{equation}
where $M_{eq}$ is given by Eq.~(6). From (6) and (11), we have $\Delta
S(t=\gamma^2/2) = 0$, i.e., $\Delta S(T)$ becomes equal to zero at the
tricritical point $A$. Besides we find from (6) and (11) that at the
triple point $C$ the entropy jump is given by
\begin{equation}
\label{eq12} \Delta S(t=0) =
  -\frac{\gamma^2}{4}\left(\frac{\alpha_s}{\sqrt{b}} + \frac{\alpha_f}{\sqrt{\beta}}
   \right)\:.
\end{equation}

On the line $BC$ the entropy jump is defined by $\Delta S_{BC}(T) =
[S_{FS}(T)-S_{FM}(T)]$. We obtain
\begin{equation}
\label{eq13} \Delta S_{BC}(T) =
 \left( |t| -\frac{\gamma^2}{4}\right)\left(\frac{\alpha_s}{\sqrt{b}}
  + \frac{\alpha_f}{\sqrt{\beta}}
   \right)\:.
\end{equation}
At the tricritical point $B$ this jump is equal to zero as it should
be. The calculation of the specific heat jump on the first order phase
transition lines $AC$ and $BC$ is redundant for two reasons. Firstly,
the jump of the specific heat at a first order phase transition differs
from the entropy by a factor of order of unity. Secondly, in caloric
experiments where the relevant quantity is the latent heat $Q = T
\Delta S(T)$, the specific heat jump can hardly be distinguished.

{\bf 5. Concluding remarks} 

The results obtained in this paper can be
used in interpretations of magnetic and caloric experiments in various
classes of magnetic superconductors with a spin-triplet Cooper pairing.
Our general consideration is irrespective of the ratio between $T_s$
and $T_f$. But one should be aware that depending on the concrete
choice of the substance, either $T_s < T_f$ or $T_s \geq T_f$ may
happen. For the new class of ferromagnetic compounds, mentioned in
Sec.~1, $T_s \ll T_f$, and this must be taken in mind  when the present
theory is compared with the experiments performed with these metallic
compounds. The condition $T_f
> T_s$ leads to a reduction of the possible phases and phase
transitions. For $T_f > T_s$, the diagram shown in Fig.~1 is reduced to
the domain in the plane ($t$, $r$) where the inequality $r >
(\alpha_s/\alpha_f) t$ is fulfilled. Experimental data for the density
of states and, hence, for the parameters $\alpha_s$ and $\alpha_f$ are
needed in order to locate the respective part of the plane ($t$, $r$).
According to the current literature~[7] the respective experimental
data for the metallic compounds enumerated in the Sec.~1 are not yet
available. It is, however, obvious that the metallic compounds with
$T_f \gg T_s$ are described by the phase transition lines in the fourth
quadrant of the ($t$, $r$) plane and, therefore, one may observe second
or first order FM-FS transitions depending on the ratios
$\alpha_s/\alpha_f$ and $T_s/T_f$.

{\bf Acknowledgments.} Financial support through SCENET (6FP-EC) and
Project 10-98/2003/JINR (Dubna) are acknowledged.

\begin{center}
{\bf REFERENCES}
\end{center}

[1] SAXENA S. S., P. AGARVAL, K. AHILAN, F. M. GROSCHE,, R. K. W.
HASELWIMMER, M. J. STEINER, E. PUGH, I. R. WALKER, S. R. JULIAN, P.
MONTHOUX, G. G. LONZARICH, A. HUXLEY, I. SHEIKIN, D. BRAITHWAITE, J.
FLOUQUET, Nature (London) {\bf 406}, 2000, 587-592. [2] PFLEIDERER C.,
M. UHLATZ, S. M. HAYDEN, R. VOLLMER, H. V. L\"OHNEYSEN, N. R. BERHOEFT,
C. G. LONZARICH, Nature (London) {\bf 412}, 2001, 58-61. [3] AOKI D.,
A. HUXLEY, E. RESSOUCHE, D. BRAITHWAITE, J. FLOUQUET, J-P. BRISON, E.
LHOTEL, C. PAULSEN, Nature (London) {\bf 413}, 2001, 613-616. [4]
BLAGOEVA E. J., G. BUSIELLO, L. DE CESARE, Y. T. MILLEV, I. RABUFFO, D.
I. UZUNOV, Phys. Rev. {\bf B42}, 1990, 6124-6137. [5] UZUNOV D. I. In:
Advances in Theoretical Physics, ed. E. Caianiello, Singapore, World
Scientific, 1990, 96-135. [6] UZUNOV D. I., Theory of Critical
Phenomena, Singapore, World Scientific, 1993, 1-452. [7] WALKER M. B.,
K. V. SAMOKHIN, Phys. Rev. Lett. {\bf 88}, 2002, 204001-204004. [8]
SHOPOVA D. V., D. I. UZUNOV,  Phys. Lett.  {\bf A 313}, 2003, 139-143.
[9] SHOPOVA D. V., D. I. UZUNOV, J. Phys. Studies, {\bf 7} No 4 (in press).

\end{document}